\documentclass[aps,prx,twocolumn,superscriptaddress,longbibliography,floatfix]{revtex4-1}

\usepackage{amsmath}
\usepackage{graphicx}
\usepackage[dvipsnames]{xcolor}
\usepackage[colorlinks,linkcolor=blue,citecolor=blue,urlcolor=blue]{hyperref}
\usepackage{stix}
\begin{document}

\newcommand{\dd}{\mathrm{d}}
\renewcommand{\vec}[1]{\mathbf{#1}}
\newcommand{\gvec}[1]{\boldsymbol{#1}}
\newcommand{\en}{\varepsilon}
\newcommand{\hc}{\hat{c}}
\newcommand{\hcd}{\hat{c}^\dagger}
\newcommand{\hd}{\hat{d}}
\newcommand{\hdd}{\hat{d}^\dagger}
\newcommand{\pp}{p_\perp}
\newcommand{\pol}{\hat{\epsilon}}
\newcommand{\nn}{\nonumber}

\mathchardef\mhyphen="2A
\newcommand{\dx}[1]{\!\mathop{\mathrm{d}}\!#1\,}
\let\OldRe\Req
\renewcommand{\Re}{\text{Re}\,}
\let\OldIm\Im
\renewcommand{\Im}{\text{Im}\,}
\newcommand{\iu}{\mathrm{i}}

\newcommand{\intd}[1]{\int\!\!\dd #1\,}
\newcommand{\iintd}[2]{\int\!\!\dd #1\!\!\int\!\!\dd #2\,}

\newcommand{\toadd}[1]{\textcolor{cyan}{[#1]}}
\newcommand{\tocheck}[1]{\textcolor{red}{#1}}
\newcommand{\cc}[2]{{\textcolor{blue}{#1}}{\textcolor{green}{[#2]}}}

\newcommand{\ph}{\mbox{$p$--$h$}}

\hyphenation{photo-emission}


\author{Michael Sch\"uler}
\affiliation{Department of Physics, University of Fribourg, 1700
  Fribourg, Switzerland}
\affiliation{Stanford Institute for Materials and Energy Sciences (SIMES),
SLAC National Accelerator Laboratory, Menlo Park, CA 94025, USA}
\author{Umberto De Giovannini}
\affiliation{Max Planck Institute for the Structure and Dynamics of
  Matter, Luruper Chaussee 149, 22761 Hamburg, Germany}
\author{Hannes H\"ubener}
\affiliation{Max Planck Institute for the Structure and Dynamics of
  Matter, Luruper Chaussee 149, 22761 Hamburg, Germany}
\author{Angel Rubio}
\affiliation{Max Planck Institute for the Structure and Dynamics of
  Matter, Luruper Chaussee 149, 22761 Hamburg, Germany}
\affiliation{Center for Computational Quantum Physics (CCQ), 
  The Flatiron Institute, 162 Fifth avenue, New York NY 10010}
\author{Michael A. Sentef}
\affiliation{Max Planck Institute for the Structure and Dynamics of
  Matter, Luruper Chaussee 149, 22761 Hamburg, Germany}
\author{Philipp Werner}
\affiliation{Department of Physics, University of Fribourg, 1700 Fribourg, Switzerland}

\title{Local Berry curvature signatures in dichroic angle-resolved photoelectron spectroscopy
}

\begin{abstract}
  Topologically nontrivial two-dimensional materials hold great promise for next-generation optoelectronic applications. However, measuring the Hall or spin-Hall response is often a
  challenge and practically limited to the ground state. An
  experimental technique for tracing the topological character in a
  differential fashion 
  would provide useful insights. In this work, we show that circular dichroism angle-resolved 
  photoelectron spectroscopy (ARPES) provides a powerful tool which can
  resolve the topological and quantum-geometrical character in momentum space.
  In particular, we investigate how to map out the
  signatures of the \emph{local} Berry curvature by exploiting its intimate connection
  to the orbital angular momentum. A spin-resolved detection of the photoelectrons allows to extend the approach to spin-Chern insulators. Our predictions are corroborated by state-of-the art \emph{ab initio} simulations employing time-dependent density functional theory, complemented with model calculations. The present proposal can be extended to address topological properties in materials  out of equilibrium in a time-resolved fashion.
\end{abstract}

\maketitle

\section{Introduction}

The discovery of the remarkable physical consequences in materials of the Berry
curvature of wave-functions has spurred progress across many
research fields in physics. In periodic solids, the most notable
examples are topological insulators (TIs) and superconductors
~\cite{hasan_colloquium:_2010, qi_topological_2011}, in which a global
topological invariant emerges from momentum-space integrals of the
Berry curvature. This global topology gives rise for example to a
quantized Hall conductance carried by surface or edge
states~\cite{hasan_colloquium:_2010}. In particular, two-dimensional
(2D) systems are currently in the spotlight, for their flexibility in
creating van der Waals heterostructures and thus potentially
next-generation transistor devices~\cite{kou_two-dimensional_2017}.
However, independently of global topology, it is becoming increasingly
evident that also local quantum geometry can have dramatic physical
consequences. Haldane pointed out the consequence of Berry curvature
on the Fermi surface for Fermi-liquid transport properties
\cite{haldane_berry_2004}, reinterpreting the Karplus-Luttinger
anomalous velocity \cite{karplus_hall_1954} in modern Berry phase
language. Similarly, a geometrical description of the fractional
quantum Hall effect was proposed
\cite{haldane_geometrical_2011}. Examples of physical consequences of
quantum geometry, expressed as the Fubini-Study metric, include
unusual current-noise characteristics \cite{neupert_measuring_2013} or
the geometric origin of superfluidity in flat-band systems
\cite{julku_geometric_2016}. Other prominent examples for the impact
of local Berry curvature are strongly anisotropic high-harmonic
generation signals from hexagonal boron nitride (hBN) or transition
metal
dichalogenides~\cite{li_probing_2013,tancogne-dejean_atomic-like_2018},
the valley Hall
effect~\cite{barre_spatial_2019,Shin:2019kt} and chiral photocurrents in topological 
Weyl semimetals~\cite{rees_quantized_2019,ma_direct_2017}.
 Also, the recently discovered nonlinear Hall effect~\cite{sodemann_quantum_2015,xu_electrically_2018,ma_observation_2019} in topologically trivial systems is an important manifestion of local Berry curvature effects.

In contrast to cold atoms in optical lattices, where measurements of local Berry curvature were  recently demonstrated \cite{flaschner_experimental_2016}, 
the observation of the local Berry curvature in materials still poses a challenge.  
Although remarkable
progress~\cite{wang_quantum_2013,reis_bismuthene_2017,li_theoretical_2018,marrazzo_prediction_2018}
in predicting and realizing large-gap 2D TIs has been made,
alternative efficient ways of exploring topological properties are necessary to further advance this active branch of materials research. Recent theoretical
proposals~\cite{tran_probing_2017,schuler_tracing_2017} and
experimental realizations~\cite{asteria_measuring_2019} in ultracold
atomic gases have demonstrated a quantization of 
circular dichroism in the
photoabsorption, which enables a clear distinction between topologically
trivial and nontrivial phases. Similarly, dichroic selection rules determine the optical absorption 
of 2D materials, especially in presence excitons~\cite{cao_unifying_2018}.
The underlying mechanism is -- similar
to magnetic systems~\cite{souza_dichroic_2008} -- the intrinsic
magnetization resulting from orbital angular momentum (OAM).
In this work, we demonstrate that the extension of this approach to
angle-resolved photoemission (ARPES) with circularly polarized light
provides direct information on the Berry curvature in 2D
systems. Unlike photoabsorption, the circular dichroism in the angular distribution
is sensitive to the momentum-resolved OAM and thus gives access
to valley-resolved topological properties. This enables tracing the
\emph{local} Berry curvature, which is hardly accessible by other
experimental techniques.

We demonstrate the connection between circular dichroism, OAM and the Berry
curvature by considering simple tight-binding (TB) models and confirm
our findings by state-of-the-art \emph{ab initio} calculations~\cite{de_giovannini_first-principles_2017} based
on real-time time-dependent density functional theory
(TDDFT)~\cite{Marques:2011ud,Ruggenthaler:2018ht}.  The latter formalism 
provides a realistic description of the full ionization process
including final-state effects, transport through material, electron-electron interaction and non-equilibrium 
dynamics~\cite{UDeGiovannini:2012hy,DeGiovannini:2013dr,de_giovannini_monitoring_2016,Hubener:2018id,Sato:2018jw,Krecinic:2018fh}.
While we will focus the discussion
on paradigmatic systems similar to graphene, our results are generic
and can be applied to other 2D materials.

\section{Berry curvature and orbital angular momentum}

OAM and the resulting orbital magnetization is a fundamental property
of the Bloch wave-functions of individual bands and has an intimate
connection to the Berry curvature.  To illustrate this relation
and its manifestation in ARPES, let us consider a generic 2D
material, possibly with spin-orbit coupling (SOC). In
strictly 2D systems, the $z$-projection of the spin $S_z$ 
is still an exact quantum number even in the
presence of SOC~\cite{li_theoretical_2018},
 such that the general
Bloch Hamiltonian in the spinor basis reads
\begin{align}
\label{eq:Ham}
  \hat{H}(\vec{k}) = \begin{pmatrix} \hat{h}_{\uparrow}(\vec{k}) & 0 \\ 0 &
    \hat{h}_{\downarrow}(\vec{k})  \end{pmatrix} \ .
\end{align}
Note that the finite spread of the binding potential 
in the out-of-plane direction as well as structural deviations from 2D geometry (such as buckling)  break, in principle, the conservation of $S_z$. Nevertheless, the 
Hamiltonian~\eqref{eq:Ham} provides an excellent approximation for the systems investigate here. The validity of this description is underpinned in Appendix~\ref{sec:soc2d}.

In the absence of magnetism, time-reversal symmetry (TRS) holds,
constraining
$\hat{h}_{\downarrow}(\vec{k}) =[\hat{h}_{\uparrow}(-\vec{k})]^*$ and
giving rise to a degeneracy of the spin-resolved bands:
$\hat{h}_\sigma(\vec{k})|u_{\vec{k}\alpha\sigma}\rangle =
\en_{\vec{k}\alpha}|u_{\vec{k}\alpha\sigma}\rangle$.

\begin{figure}[t]
  \includegraphics[width=0.9\columnwidth]{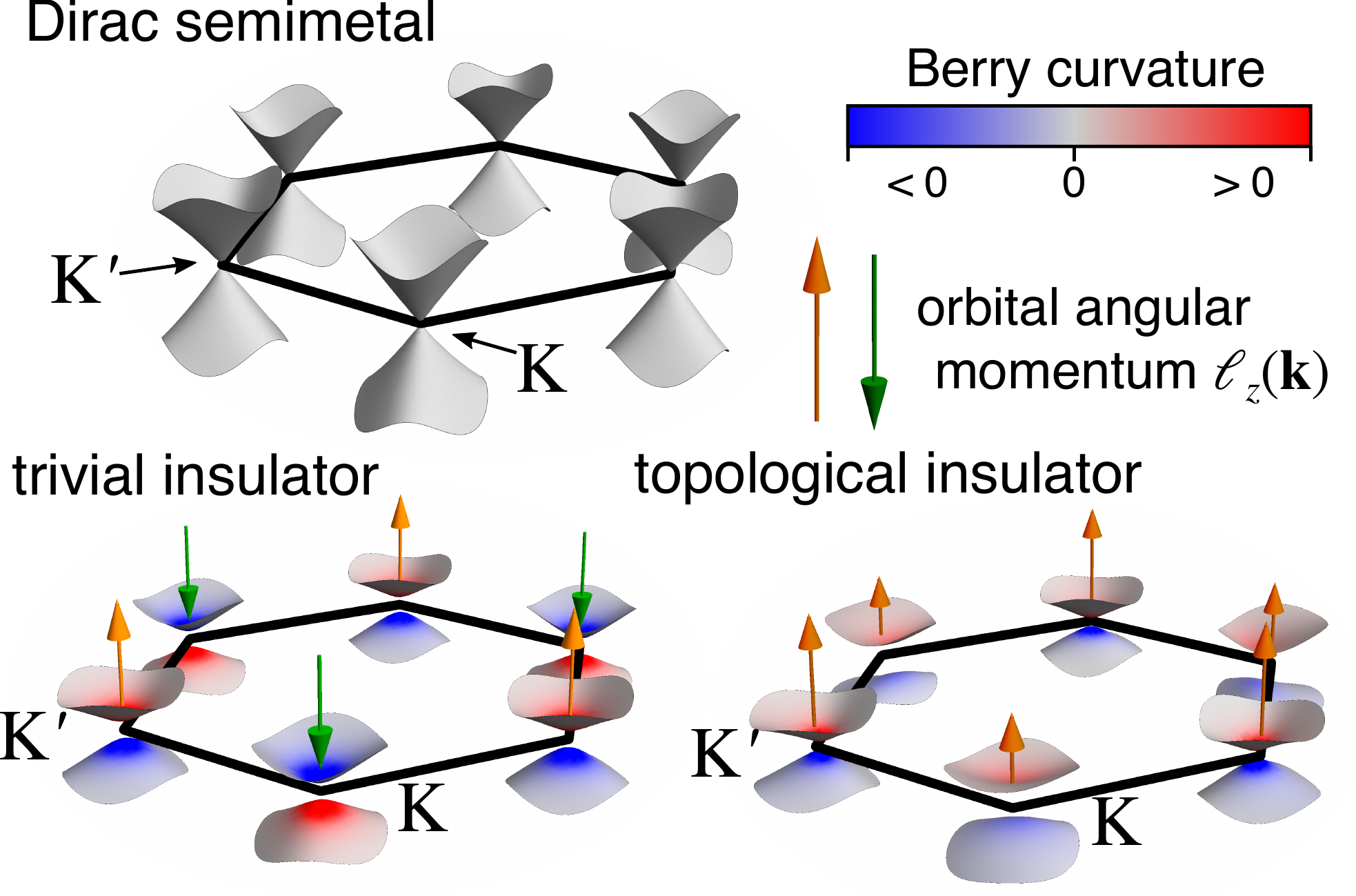}
  \caption{Illustration of the Berry curvature and orbital angular momentum. The surfaces 
  represent the valence and conduction band, while the coloring indicates the Berry curvature.
  The arrows illustrate the orbital angular momentum of the valence band. \label{fig:oam_bz_illustration} }
\end{figure}


The individual bands possess intrinsic properties which are determined
by the band structure topology. An important example of such properties is the
OAM. The full description of the OAM and the related orbital
magnetization in terms of the Berry phase
theory~\cite{thonhauser_orbital_2005,xiao_berry_2005} (so-called
modern theory of polarization) has been formulated relatively recently. For a band
$\alpha$ with spin $\sigma$, the orbital moment is defined as
\begin{align}
  \label{eq:oam}
  \ell_z(\vec{k})  = \frac{m}{\hbar}  \mathrm{Im}\langle
  \partial_{k_x} u_{\vec{k} \alpha\sigma} | \hat{h}_{\sigma}(\vec{k}) -
  \en_{\vec{k}\alpha} |  \partial_{k_y} u_{\vec{k} \alpha\sigma} \rangle .
\end{align}
This shows that the orbital magnetization
$m_z(\vec{k}) = (e/m) \ell_z(\vec{k})$ is an \emph{intrinsic} property
of the underlying band, related to self-rotation, which can emerge even
if no magnetic atoms are present.

The orbital moment~\eqref{eq:oam} transforms exactly as the Berry
curvature under symmetry operations, underpinning their tight
connection.  In particular, in the case where the Berry curvature is primarily due
to a valence ($v$) band coupling to a conduction ($c$) band, the OAM
becomes proportional to the Berry curvature
$\Omega_{v\sigma}(\vec{k})$ of the valence band:
\begin{align}
  \label{eq:oam_berry}
  \ell_z(\vec{k}) = -\frac{m}{\hbar} (\en_{\vec{k}c}- \en_{\vec{k}v})
  \Omega_{v\sigma}(\vec{k}) \ .
\end{align}
TRS implies $\Omega_{v\uparrow}(\vec{k})
= -\Omega_{v\downarrow}(-\vec{k})$, while inversion symmetry results
in $\Omega_{v\uparrow}(\vec{k})
= \Omega_{v\uparrow}(-\vec{k})$. Hence, in systems possessing both
symmetries, $\Omega_{v\uparrow}(\vec{k})=-\Omega_{v\downarrow}(\vec{k})$ holds; 
in absence of SOC the Berry curvature and thus the OAM vanishes
exactly. 
Therefore, measuring the momentum-resolved OAM allows to
map out the local Berry curvature.

For graphene-like insulating systems, the Berry
curvature and the OAM for the
three possible scenarios are sketched in
Fig.~\ref{fig:oam_bz_illustration}. Graphene (neglecting the SOC)
possesses inversion symmetry, and respective sublattice sites on the
honeycomb lattice are equivalent; hence $\ell_z(\vec{k})$ is zero in
both spin channels, giving rise to a Dirac semimetal. Breaking
inversion symmetry -- for instance by considering systems with
inequivalent atoms on the respective sublattice sites as in hexagonal
boron nitride (hBN) -- opens a gap and generates a nonzero Berry curvature. 
The resulting trivial insulator shows OAM at the Dirac points
K and K$^\prime$ with opposite sign due to TRS. The system is
characterized by a nonzero valley Chern
number $C_\mathrm{val}(K)= -C_\mathrm{val}(K^\prime) =
\int_{\mathrm{val}(K)}d \vec{k}\, \Omega_{v}(\vec{k})/2\pi$, indicating 
a pronounced valley magnetization~\cite{gunawan_valley_2006}.

Spin-orbit coupling in graphene-like systems 
renders them (type-II) $\mathbb{Z}_2$ spin Chern insulators \cite{hasan_colloquium:_2010}, according to the
Kane-Mele mechanism~\cite{kane_$z_2$_2005}. The bands exhibit an inverted orbital
character at K and K$^\prime$, respectively, while the TRS is broken for
each spin channel individually (even though the system possesses global
TRS). Considering the total OAM, the spin Chern number $C_s=\pm 1$ indicates a total
\emph{chiral} $L_z = \int d \vec{k}\, \ell_z(\vec{k}) \ne 0$, with
the same magnitude and opposite sign for spin-up and spin-down
electrons, respectively.

While optical techniques sensitive to a total chirality -- such as
magnetic circular dichroism (MCD) -- cannot separate out the individual
spin channels, advances in spin-resolved
ARPES~\cite{okuda_recent_2017} (sARPES) enable a selective measurement
of spin-up or spin-down photoelectrons. A
dichroic sARPES measurement would allow to map out momentum-
and spin-resolved OAM properties, which is hard to achieve by other methods.
Recent experiments on chiral surface states in TIs~\cite{park_chiral_2012} 
demonstrate the feasibility of detecting circular dichroism in photoemission.

\section{Circular dichroism in spin- and angle-resolved photoemission}

To discuss how the OAM is reflected in ARPES we consider the photoemission intensity as 
described by Fermi's Golden Rule in the dipole approximation~\cite{schattke_solid-state_2008}
\begin{align}
  \label{eq:arpes}
  I(\vec{p},\en_f) \propto \left|\langle \chi_{\vec{p},p_\perp} |
  \hat{\epsilon}\cdot \hat{D} | \psi_{\vec{k}\alpha}\rangle\right|^2
  \delta(\en_{\vec{k}\alpha} + \hbar \omega - \en_f) \ ,
\end{align}
where $|\psi_{\vec{k}\alpha}\rangle$ denotes the Bloch state
corresponding to the cell-periodic wave-function
$|u_{\vec{k}\alpha}\rangle$. The photon energy is given by $\hbar
\omega$, and $\en_f=(\vec{p}^2+p^2_\perp)/2$ is the energy of the 
photoelectron final state $|\chi_{\vec{p},p_\perp}\rangle$. The matrix
element of the dipole operator $\hat{D}$ and the
polarization direction $\hat{\epsilon}$ determine the selection
rules. The in-plane momentum $\vec{p}$ is identical to the
quasi-momentum $\vec{k}$ up to a reciprocal lattice vector. We can
extend Eq. ~\eqref{eq:arpes} to the spin-resolved
intensity $I_{\sigma}(\vec{p},\en_f)$ by assuming a spin-resolved detection of 
the final states $|\chi_{\vec{p},p_\perp,\sigma}\rangle$, fixing the
photoelectron spin $\sigma$.

\begin{figure*}[t]
  \includegraphics[width=\textwidth]{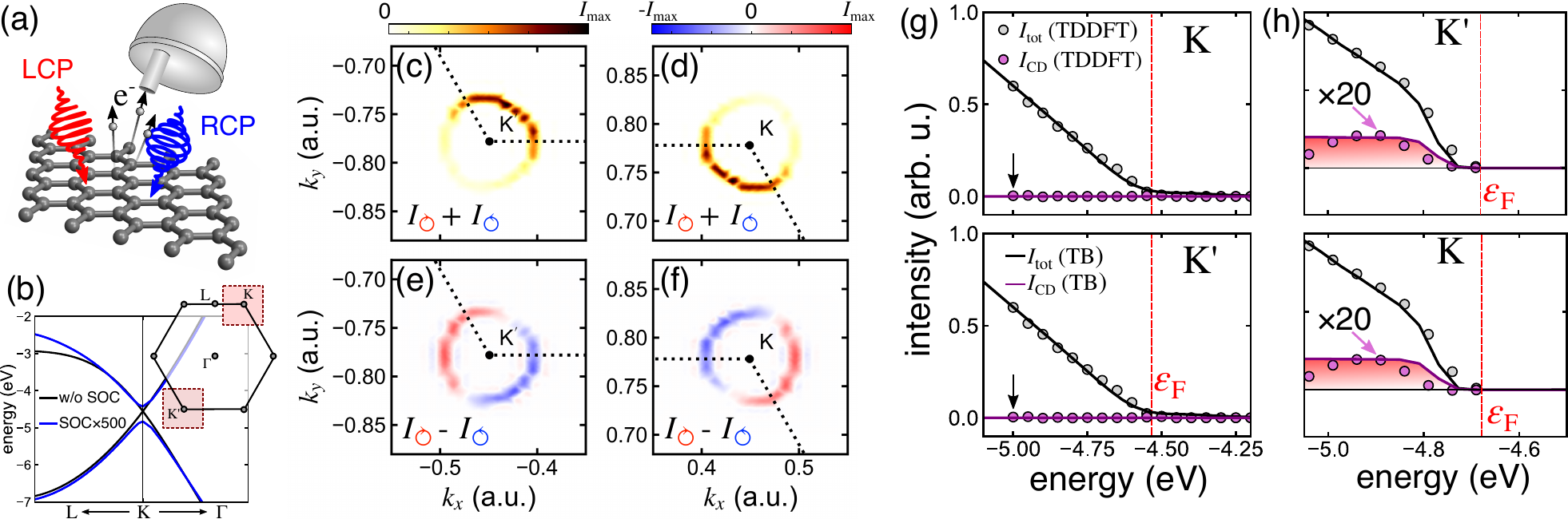}
  \caption{(a) Sketch of the calculation setup: photoemission by left- (LCP) or right- (RCP) hand
    circularly polarized light, with polarization vector in the plane.
  (b) Band structure of graphene close to the Dirac point with zero (black line) and enhanced (blue line) SOC, respectively, obtained from density functional theory (DFT). The first Brillouin zone (BZ) of the system with honeycomb lattice is shown in the inset. The
  shaded boxes indicate the magnified regions shown in (c)--(f). (c) and
  (d): total ARPES intensity $I_\mathrm{tot}(\vec{k},\en_f)$ (normalized to its maximum value $I_\mathrm{max}$) at
  $\en_f=47$\,eV, close to the K and K$^\prime$ point,
  respectively. (e)--(f): corresponding dichroic signal
  $I_\mathrm{CD}(\vec{k},\en_f)$. (g): Integrated signal (over the
  shaded regions in (b))
  $I_\mathrm{tot}$ and  $I_\mathrm{CD}$ as a function of the binding
  energy. The black arrow indicates the energy for which the
  angle-resolved intensities in (c)--(f) are shown. (h) Integrated signal from spin-up electrons (analogous to (g)) for graphene with enhanced SOC. \label{fig:gra} }
\end{figure*}

To detect OAM textures, we exploit the circular dichroism in ARPES. As sketched in
Fig.~\ref{fig:gra}(a) we consider the experimental situation, 
where the probe field is either left-hand (LCP) or
right-hand circularly polarized (RCP), with the 
polarization vector $\hat{\epsilon}^{(\pm)}$ in plane, i.\,e. normal incident fields. 
The corresponding ARPES intensities $I^{(\pm)}(\vec{k},\en_f)$
then define the total (unpolarized) 
$I_\mathrm{tot}(\vec{k},\en_f)=I^{(-)}(\vec{k},\en_f)+I^{(+)}(\vec{k},\en_f)$ and circular dichroism
$I_\mathrm{CD}(\vec{k},\en_f)=I^{(-)}(\vec{k},\en_f)-I^{(+)}(\vec{k},\en_f)$ signal.

\subsection{Connection to orbital angular momentum}

The close connection between the dichroic signal $I_\mathrm{CD}(\vec{k},\en_f)$ and the Berry curvature is already apparent from symmetry considerations. TRS dictates 
$I_\mathrm{CD} (\vec{k},\en_f)= -I_\mathrm{CD}(-\vec{k},\en_f)$, such that the circular dichroism
integrated over the whole Brillouin zone (BZ) vanishes. In addition, a system
possessing inversion symmetry results in an exactly vanishing valley-integrated circular dichroism, analogous to the Berry curvature. This argument demonstrates that the breaking of
TRS -- a characteristic property of Chern insulators -- is reflected in a nonzero
\emph{total} circular dichroism.

The manifestation of local OAM chirality in the circular dichroism can be understood intuitively in terms of the wave packet picture, which is also playing a fundamental role in the theory of orbital magnetization~\cite{xiao_berry_2010}. Instead of a Bloch initial state $|\psi_{\vec{k}_c\alpha}\rangle$, we can consider a wave packet $|W_{\vec{k}\alpha}\rangle$ composed of momenta close to $\vec{k}$. 
Hence, $W_{\vec{k}\alpha}(\vec{r})$
has a finite spread in real space and 
properties similar 
to a molecular orbital. In particular, its OAM is given by 
$\langle \hat{L}_z \rangle = \langle W_{\vec{k}\alpha} | \hat{L}_z | W_{\vec{k}\alpha} \rangle$; 
in the limit of an infinitely sharp distribution, such that $|W_{\vec{k}\alpha}\rangle$ becomes identical to $|\psi_{\vec{k}\alpha}\rangle$, one finds $\langle \hat{L}_z \rangle =\ell_z(\vec{k})$. Nonzero $\langle \hat{L}_z \rangle$ indicates self-rotation of the wave packet, which will be reflected in the dipole selection rules in the ARPES matrix elements in Eq. ~\eqref{eq:arpes}. 

This picture can be used to obtain a qualitative description of the dichroism, as detailed in Appendix~\ref{sec:wp_details}. Introducing the analogue of a cell-periodic function by $F_{\vec{k}\alpha}(\vec{r}) = e^{-\iu\vec{k}\cdot\vec{r}} W_{\vec{k}\alpha}(\vec{r})$, its OAM properties can be analyzed by projecting it onto eigenfunctions of $\hat{L}_z$: $F_{\vec{k}\alpha}(\vec{r}) = 1/\sqrt{2\pi}\sum_m e^{\iu m \theta} \mathcal{F}_{\vec{k}\alpha,m}(s,z)$, where $(s,\theta)$ are the in-plane polar coordinates. Replacing the final states by plane waves (PWs), one can approximate the matrix elements in Eq. ~\eqref{eq:arpes} by
\begin{align}
\label{eq:wp_matel}
M^{(\pm)}_\alpha(\vec{k},p_\perp) \approx \int^\infty_0\!d s\,
  \int^\infty_{-\infty}\! d z\, e^{-\iu p_\perp z} s^2 \mathcal{F}_{\vec{k}\alpha,\mp 1}(s,z) \ .
\end{align}
Therefore, the OAM properties of the initial state are directly reflected in the circular dichroism. In particular, $\langle \hat{L}_z \rangle = 0$ typically implies $\mathcal{F}_{\vec{k}\alpha,+1}(s,z)=\mathcal{F}_{\vec{k}\alpha,-1}(s,z)$; hence the circular dichroism vanishes. In Appendix~\ref{sec:wp_details}, we discuss the illustrative example of hBN and analyze the OAM properties in detail.

\subsection{Calculation of photoemission spectra}

To compute ARPES from first principles one does not need to resort to the approximated one-step model of 
photoemission like the one of Eq.~\eqref{eq:arpes}.
Instead of using Eq.~\eqref{eq:arpes} we employ TDDFT~\cite{Marques:2011ud} with the t-SURFFP
method~\cite{de_giovannini_first-principles_2017}
which avoids any reference to explicit final states by directly computing the momentum and 
energy distribution of the photocurrent created by a specific pulse field and thus
allows to compute the intensity directly from the real-time evolution. 

While the first-principles approach provides results in excellent agreement with
experiments (see below), a more intuitive understanding can be gained by
considering a simple model for the direct evaluation of Eq. ~\eqref{eq:arpes}. 
The Bloch states are represented by a TB model of atomic orbitals, while
replacing the final states $|\chi_{\vec{p},p_\perp}\rangle$ in this equation by PWs
 $|\vec{p},p_\perp\rangle$ eliminates scattering of the
photoelectron from the lattice, which allows us to focus on the intrinsic
contribution of the Bloch states to the ARPES intensity. The matrix elements 
in Eq. ~\eqref{eq:arpes} are computed in the length gauge, which encodes the 
selection rules with respect to the LCP or RCP polarization. In what follows, we refer to the
resulting model as TB+PW model.
Furthermore, an analytical treatment is possible in certain cases, 
providing a clear physical picture. 

The combination of 
the two methods allows a comprehensive analysis. The details on both methods can be found in Appendix~\ref{sec:tddft_details} and \ref{sec:tb_details}.

\section{Results}

Here we investigate the three classes of 2D materials represented in Fig.~\ref{fig:oam_bz_illustration}, namely the Dirac semimetal, trivial insulator and topological insulator, and identify the distinct features of the circular dichroism in ARPES. 
The Dirac semimetal we consider is graphene, while hBN exemplifies a trivial insulator. As examples of topological insulators we study bismuthane and graphene with artificially enhanced SOC.

%

\begin{figure}[t]
  \centering
  \includegraphics[width=0.9\columnwidth]{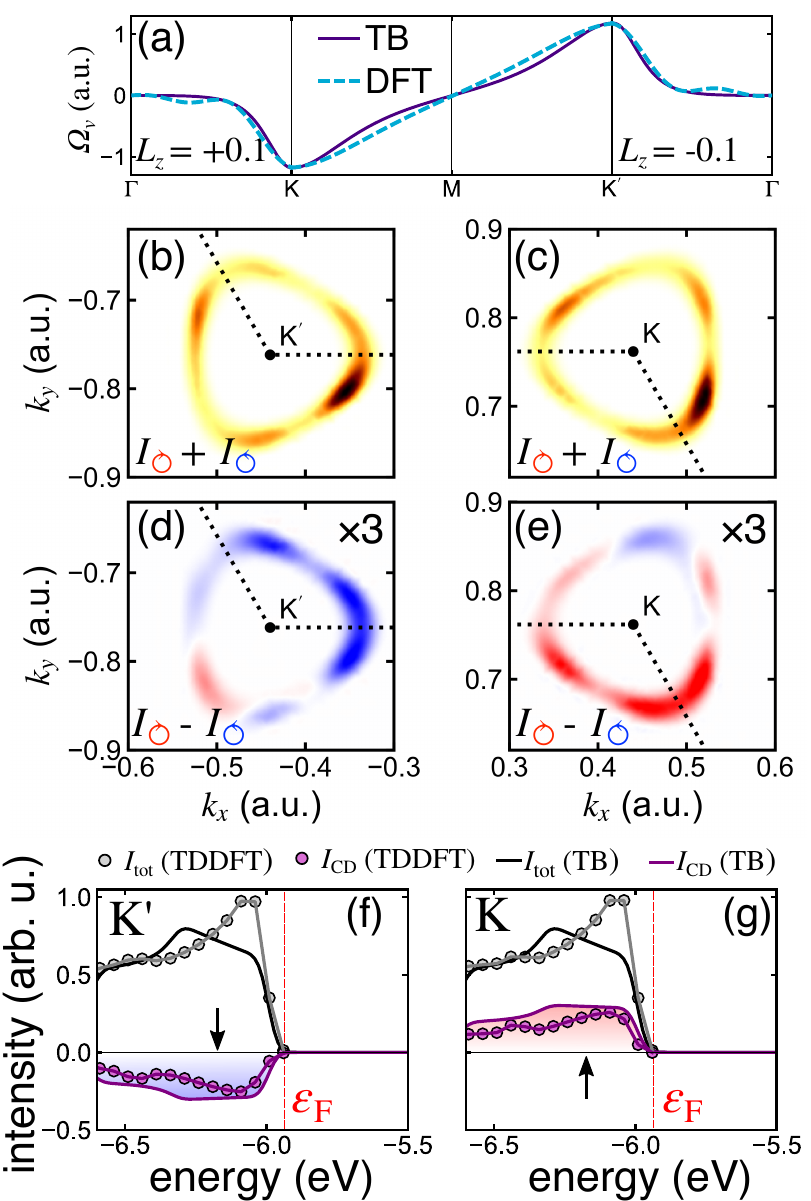}
  \caption{(a) Berry curvature $\Omega_v(\vec{k})$ of the top valence
    band, comparing TDDFT and TB
    result. The OAM amounts to $L_z=0.1$ at K and $L_z=-0.1$ at K$^\prime$.
    (b)--(c): total ARPES intensity $I_\mathrm{tot}(\vec{k},\en_f)$ at
    $\en_f=45.811$\,eV. (d)--(e): Corresponding circular dichroism. The color coding is analogous to Fig.~\ref{fig:gra}. The integrated
    signals as a function of the binding energy at K$^\prime$ (f) and K
    (g), respectively.  \label{fig:hbn}
    }
\end{figure}

\subsection{Graphene}

We start by discussing ARPES from
graphene, which is the prototype of a 2D material. We focus on the regions in the
first BZ close
to the two inequivalent Dirac points (Fig.~\ref{fig:gra}(b)). The
photon energy is fixed at $\hbar \omega=52$\,eV. Neglecting the very weak SOC, 
spin resolution is not required at this point.

Figure~\ref{fig:gra}(c)--(d) shows a typical ARPES cut at fixed $\en_f$,
obtained by the t-SURFFP approach. Consistent with
experiments~\cite{mucha-kruczynski_characterization_2008,hwang_direct_2011},
the prominent dark corridor (region of minimal intensity) is observed
in the $\Gamma$--K or $\Gamma$--K$^\prime$ direction at this photon
energy. The dark corridor is a consequence of destructive 
interference of the emission from the two sublattice
sites~\cite{hwang_direct_2011}, which can be illuminated by
$s$-polarized light~\cite{gierz_illuminating_2011}.

The calculated dichroic signal
$I_\mathrm{CD}(\vec{k},\en_f)$, 
shown in Fig.~\ref{fig:gra}(e)--(f), is in very good agreement with 
experimental data reported in Ref.~\cite{gierz_graphene_2012}. In particular, 
when following a path perpendicular to the $\Gamma$--K
direction, the chiral character is consistent with the experimental data from Ref.~\cite{liu_visualizing_2011}. As is apparent from
Fig.~\ref{fig:gra}(e)--(f), the valley-integrated circular dichroism vanishes. This is
confirmed by both theoretical methods, shown in Fig.~\ref{fig:gra}(g), where we compare the integrated TDDFT results
to those of the TB+PW model 
for an integration range corresponding to the two shaded
regions in Fig.~\ref{fig:gra}(b), finding excellent agreement. Hence, the dichroic properties provide a direct proof of the vanishing Berry curvature.

\subsection{Hexagonal boron nitride}

We now turn to the paradigmatic case of a trivial insulator with broken inversion symmetry (as sketched in Fig.~\ref{fig:oam_bz_illustration}) 
by studying single-layer hBN. Similar to graphene, hBN is 
a $\pi$-conjugated system dominated by $p_z$ orbitals on the sublattice
sites with a large ionic-like band gap. The Berry curvature becomes finite and
very pronounced around the K and K$^\prime$ points.
Comparing $\Omega_v(\vec{k})$ of the top valence band within the TB
model and the first-principles calculation (Fig.~\ref{fig:hbn}(a)), the
excellent agreement indicates that the orbital mixing of the top
valence and bottom conduction band -- of predominant $p_z$
orbital character -- gives the main contribution to
$\Omega_v(\vec{k})$. Hence, Berry curvature and OAM $\ell_z(\vec{k})$ are
are proportional to each other, c.\,f. Eq. ~\eqref{eq:oam_berry}. The valley-integrated OAM is $L_z\approx \pm 0.1$ with opposite sign at K and K$^\prime$,
respectively (Fig.~\ref{fig:hbn}(a)). We have also explictly evaluated wave packets and the associated OAM in Appendix~\ref{sec:wp_details}. The prediction of the dichroism from Eq. ~\ref{eq:wp_matel} is qualitatively in line with the calculated circular dichroism in Fig.~\ref{fig:hbn}(d)--(e). 

The valley-resolved measurement provided by ARPES -- as opposed to MCD~\cite{souza_dichroic_2008} -- allows to trace the valley
OAM~\cite{yao_valley-dependent_2008}. Because of the direct link to the local Berry curvature (Eq. ~\eqref{eq:oam_berry}) this provides a way of measuring the valley Chern number.

Figure~\ref{fig:hbn}(b)--(c) shows the unpolarized signal
$I_\mathrm{tot}(\vec{k},\en_f)$ for hBN close to the K and K$^\prime$
points. Note the suppression of the dark corridor, which
is due to the incomplete destructive interference. 
The corresponding circular dichroism
(Fig.~\ref{fig:hbn}(d)--(e)) shows -- following the behavior of
$\Omega_v(\vec{k})$ and the OAM -- an opposite character at the two
inequivalent Dirac points. While irradiating with LCP light results in
a much larger probability of creating a photoelectron in the
vicinity of the K point, RCP light dominates the emission from the
region around K$^\prime$. Integrating the momentum-resolved signals
yields a clear picture (Fig.~\ref{fig:hbn}(f)--(g)). 
The first-principles TDDFT results are qualitatively well 
reproduced by the TB+PW model, underpinning the intrinsic character of the dichroism.

\subsection{Bismuthane}

\begin{figure}[t]
  \centering
  \includegraphics[width=\columnwidth]{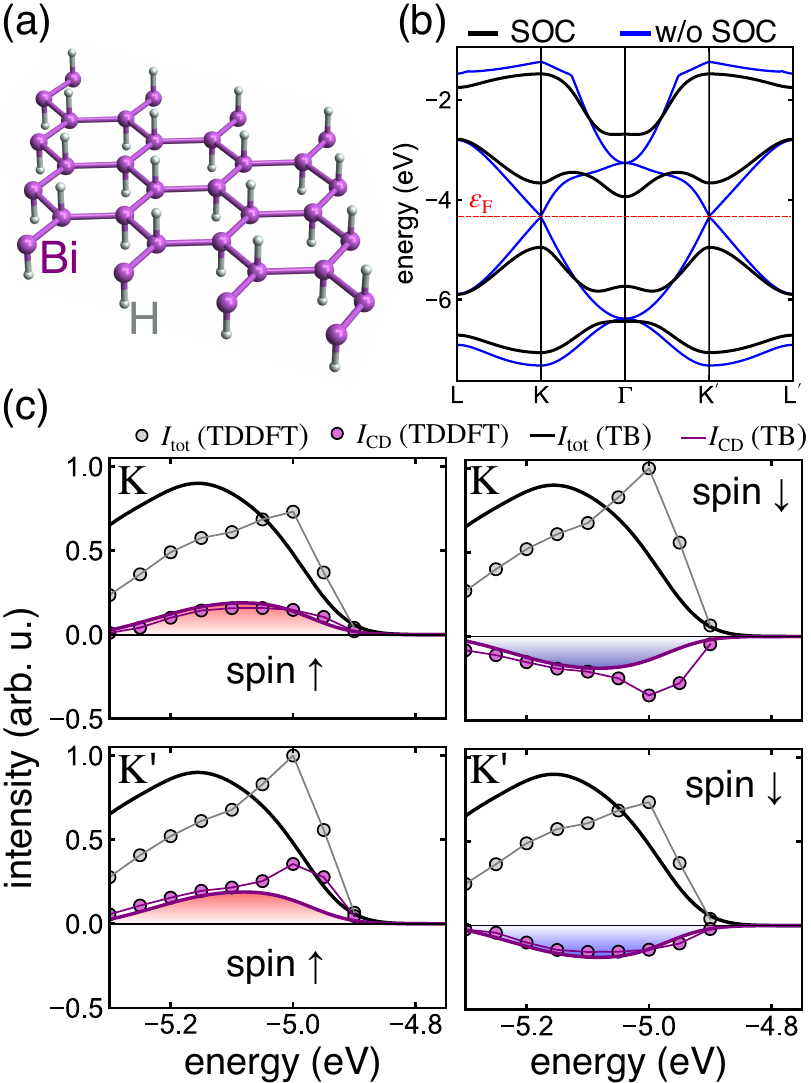}
  \caption{(a) Lattice structure of hexagonal BiH. (b) Band structure
  of BiH (obtained from DFT), fully including SOC (black) and to comparing to the case without any SOC (blue lines). (c) Integrated dichroic signal as in Fig.~\ref{fig:gra} for spin-up (left) and spin-down
  electrons (right panels) at K (upper) and K$^\prime$ (lower panels), respectively.
  \label{fig:bih}
  }
\end{figure}

To demonstrate the generic character of the connection between the Berry curvature and circular dichroism, we consider 
single-layer hydrogenated bismuthane (BiH), see Fig.~\ref{fig:bih}(a). Bismuth on the hexagonal lattice is one of most promising candidates for realizing 2D TIs~\cite{li_theoretical_2018,reis_bismuthene_2017} due to its strong intrinsic SOC.
A monolayer of hexagonal Bismuth has been experimentally characterized on a SiC substrate \cite{reis_bismuthene_2017}.
Free-standing bismuth has $p_x$, $p_y$ and $p_z$ orbitals contributing to the bands close to the Fermi energy; removing the $p_z$ orbitals from this energy range has been identified as a key mechanism~\cite{li_theoretical_2018}. 
The hydrogen atoms fulfill exactly this purpose. The system is slightly buckled, but still possesses inversion symmetry, such that the spin states are degenerate.
Artificially turning off the SOC, turns BiH into a Dirac semimetal (Fig.~\ref{fig:bih}(b)), 
while the SOC opens a large gap of $\simeq 800$~meV at K and $K^\prime$.

Due to the TRS, the Berry curvature (see Fig.~\ref{fig:oam_bz_illustration}) is opposite for spin-up and spin-down electrons, respectively. Hence, sAPRES is required to distinguish the spin species. Fig.~\ref{fig:bih}(c) shows the integrated ARPES signals for both spin channels, in analogy to the non-spin resolved case of Fig.~\ref{fig:hbn}(f) and (g). 
We are focusing on the top valence band. As expected from the case of the TI in
Fig.~\ref{fig:oam_bz_illustration}, the Berry curvature has the
sign at both K and K$^\prime$, and so has the OAM. The behavior is opposite for spin-up and spin-down, respectively; note that the global TRS implies $I_{\mathrm{CD},\uparrow}(\vec{k},\en_f)=-I_{\mathrm{CD},\downarrow}(-\vec{k},\en_f)$. Hence, the integrated circular dichroism has the same sign, confirming that BiH is a spin Chern insulator. To corroborate the topological nature of the dichroism, we have switched off the SOC within the TB+PW model.
We find vanishing valley-integrated dichroism, which is consistent with the signatures of a Dirac semimetal like graphene.

As a second example of a spin Chern insulator we can consider graphene. Even though SOC is very weak in graphene, it theoretically also renders
graphene a spin Chern insulator~\cite{hasan_colloquium:_2010}, so it is instructive in this context to study graphene with SOC. However, the SOC induced
gap of $\sim 25$\,$\mu$eV~\cite{konschuh_tight-binding_2010} is very small, so that graphene in practice 
behaves like a trivial material, as discussed above. In order to reveal the dichroic signature of the 
topologically nontrivial phase, we artificially enhance the SOC by a factor of 500. This allows to directly observe the impact of the Kane-Mele mechanism~\cite{kane_quantum_2005} on the circular dichroism. The opening of the topological gap is shown in Fig.~\ref{fig:gra}(b).
The integrated intensities in Fig.~\ref{fig:gra}(b) show a very
good agreement between the full TDDFT calculations and the TB+PW calculations for
the unpolarized intensity $I_\mathrm{tot}$. 
The circular dichroism
is overestimated by the TB+PW model by a factor of $\sim 20$. This
indicates that the circular dichroism due to scattering effects (which are missing in the TB+PW
model) is competing with the intrinsic dichroism. 
Nevertheless, the
qualitative behavior in both approaches clearly shows a non-vanishing
total circular dichroism -- and thus reveals a topologically nontrivial state.

\subsection{Universal phase diagram for graphene-like systems}

The examples for the three cases of $\pi$-conjugate systems discussed above -- Dirac semimetal (graphene), trivial insulator (hBN) and 
topological insulator (graphene with SOC) -- can all be described on the TB level by the Haldane
model~\cite{haldane_model_1988}. The Haldane model is characterized by the 
gap parameter $M$, nearest-neighbor hopping $J$, next-nearest
neighbor hopping $J^\prime$, and the associated phase $\phi$. For
$M=2.14 J$, $J^\prime=0$, $\phi=0$, the TB model of hBN is recovered,
while $M=0$, $J^\prime=-0.0473 J$,
$\phi=\mathrm{arg}(J^\prime-i \lambda)$ corresponds to the TB model
for graphene with SOC strength
$\lambda$~\cite{kane_quantum_2005,wang_quantum_2013}.

\begin{figure}[t]
  \centering
  \includegraphics[width=\columnwidth]{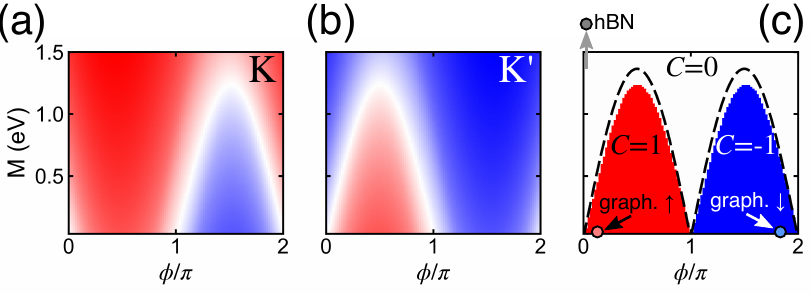}
  \caption{Energy- and valley-integrated circular dichroism $S_\mathrm{CD}$ around
    the K (a) and K$^\prime$ (b) valley, respectively, as a function
    of the phase $\phi$ and gap parameter $M$. (c):
    asymmetry signal $\Delta S_\mathrm{CD}$ (see text): red color
    corresponds to $\Delta S_\mathrm{CD}=1$, blue to $\Delta
    S_\mathrm{CD}=-1$, and white to $\Delta S_\mathrm{CD}=0$. The
    dashed lines represent the critical gap $M_\mathrm{crit}(\phi)$,
    for which $M>M_\mathrm{crit}(\phi)$ turns the system into a
    trivial insulator. The Chern numbers $C$ are given for each phase.
    The parameters are analogous to graphene, with
  $J^\prime=0.1 J$.  \label{fig:haldane}}
\end{figure}

The good qualitative agreement between the {\it ab initio} TDDFT data and the results from the TB+PW model in all considered
cases demonstrates the predictive power of the simplified description. Hence, the TB+PW approach may be used to explore
the full phase diagram of the Haldane model, providing a comprehensive
picture of the circular dichroism in graphene-like systems. 
For our analysis, we have adopted the parameters and atomic orbitals from
graphene, but replaced the TB Hamiltonian by the Haldane model. We
have computed the $\vec{k}$-integrated (over the region shown in
Fig.~\ref{fig:gra}(b)) signal $I_\mathrm{CD}(\mathrm{K}^{(\prime)},\en)$, as in
Fig.~\ref{fig:gra}, and in addition integrated over 
the binding energy $\en$. The such integrated (but valley-resolved) dichroic signal
$S_\mathrm{CD}(\mathrm{K}^{(\prime)})=\int \! d\en\, I_\mathrm{CD}(\mathrm{K}^{(\prime)},\en)$ is shown in Fig.~\ref{fig:haldane} (a)--(b). 

As Fig.~\ref{fig:haldane}(a)--(b) demonstrates, the system exhibits
a total dichroism (dominated by LCP light) for $\phi<\pi$ and small enough
$M$. For larger $M$, the dichroism stays positive around the K point, while
it becomes negative at K$^\prime$. The behavior for $\phi>\pi$ is
inverted. This suggests the following measurement strategy: if both
$S_\mathrm{CD}(\mathrm{K}) > 0$ and $S_\mathrm{CD}(\mathrm{K}^\prime) > 0$, the system
represents a Chern insulator. Similarly, $S_\mathrm{CD}(\mathrm{K}) < 0$ and
$S_\mathrm{CD}(\mathrm{K}^\prime) < 0$ should correspond to a Chern insulator
with opposite Chern number. The case $S_\mathrm{CD}(\mathrm{K})
S_\mathrm{CD}(\mathrm{K}^\prime)<0$ indicates a topologically trivial
phase. All these cases can be captured by defining $\Delta
S_\mathrm{CD} = (\mathrm{sign}[S_\mathrm{CD}(\mathrm{K})] +
\mathrm{sign}[S_\mathrm{CD}(\mathrm{K}^\prime)])/2 $, which is presented
in Fig.~\ref{fig:haldane}(c) and compared to the topological phase diagram
of the Haldane model.

Fig.~\ref{fig:haldane}(c) demonstrates a close relation between the circular dichroism and 
the topological state, since the region $\Delta S_\mathrm{CD}=1$
($\Delta S_\mathrm{CD}=-1$) is almost identical to the parameter space
with Chern number $C=1$ ($C=-1$). In contrast, the topologically
trivial regime ($C=0$) is characterized by $\Delta
S_\mathrm{CD}=0$. The corresponding topological phase diagram shown 
in Fig.~\ref{fig:haldane}(c) is also consistent with the previous
results: the hBN case would be recovered for large enough $M$ (outside
the plotted range), while the TB model for graphene with SOC for the
spin-up (spin-down) species is
equivalent to the Haldane model with $\phi=17^\circ$
($\phi=343^\circ$). 

The good agreement between the properties of the CD and the
topological phase diagram can be further supported by an analytical
evaluation of the TB+PW model (detailed in Appendix~\ref{sec:pseudospin}). 
Assuming unperturbed atomic orbitals,
we explictly calculate the matrix elements $M^{(\pm)}(\vec{k},p_\perp)=\langle
\vec{k},p_\perp|\hat{\epsilon}^{(\pm)}\cdot \vec{r}|\psi_{\vec{k}v}\rangle$
and the asymmetry $\Delta \mathcal{M}(\vec{k},p_\perp) =
|M^{(+)}(\vec{k},p_\perp)|^2-|M^{(-)}(\vec{k},p_\perp)|^2$. This
quantity is, up to the energy conservation in Eq. ~\eqref{eq:arpes},
equivalent to the dichroic ARPES intensity. Under these assumptions one can derive
\begin{align}
  \label{eq:cdasym}
  \Delta \mathcal{M}(\vec{k},p_\perp) \propto \sigma_z(\vec{k})
  [\vec{k}\times\gvec{\tau}]_z\,  \widetilde{\varphi}(k, p_\perp)
  \partial_k \widetilde{\varphi}(k, p_\perp) \ ,
\end{align}
where $\gvec{\tau}$ is the vector connecting the sublattice sites, while
$\widetilde{\varphi}(k, p_\perp)$ is the Fourier transform of the atomic $p_z$
wave function (depending on the modulus $k=|\vec{k}|$ only).

The most important term is the orbital pseudospin
$\sigma_z(\vec{k})=P_\mathrm{A}(\vec{k})-P_\mathrm{B}(\vec{k})$,
measuring the difference in orbital occupation
$P_\mathrm{A,B}(\vec{k})$ of the sublattice sites. In a topologically
trivial state, only the lower-energy site is predominantly occupied
(for instance, the nitrogen site in hBN), hence
$\sigma_z(\vec{k}) < 0$ across the whole BZ. Therefore,
Eq. ~\eqref{eq:cdasym} yields opposite signs at
$\vec{k}=\mathrm{K}$ and $\vec{k}=\mathrm{K}^\prime$.  In contrast,
in a topologically nontrivial state the orbital inversion leads to a
change of sign of $\sigma_z(\vec{k})$ in the BZ. In particular,
$\sigma_z(\mathrm{K})$ and $\sigma_z(\mathrm{K}^\prime)$ must have
opposite signs. Therefore, the asymmetry~\eqref{eq:cdasym} has the
same sign at both K and K$^\prime$. Hence, the analytical model
clearly shows that the total dichroism changes at a topological phase
transition.

\section{Discussion and Conclusion}

We have presented a detailed investigation of ARPES and, in
particular, the circular dichroism from 2D graphene-like systems. The results were
obtained by first-principles calculations of the ARPES intensity based
on TDDFT, and complemented by the analysis of a simple TB model.

In general, circular dichroism in photoemission can have multiple origins. For
instance, interference and scattering effects from the lattice 
give rise to distinct dichroism. However, in a system
possessing both inversion and TRS (like graphene without SOC), the valley-integrated CD
vanishes. Our main focus was not the dichroism related to lattice effects, but
that 
originating from 
an intrinsic property of the underlying band. In this
context, hBN is an ideal test system. In this case, the broken inversion symmetry
gives rise to a pronounced Berry curvature and associated OAM. The
distinct OAM states at the two inequivalent Dirac points directly
translates into a pronounced valley-integrated dichroism. This is underpinned by
the TB model. We stress that the connection between the OAM and the dichroism is
generic and not restricted to graphene-like systems, as confirmed by 
the example of single-layer BiH. A pronounced valley dichroism can also be expected
in monolayer transition
metal dichalcogenides of the type MX$_2$~\cite{Bertoni:2016eo}. Hence, dichroic ARPES provides an
excellent tool for studying the OAM and valley topological effects. We stress 
that the sensitivity to the local Berry curvature is a distinct feature of (spin-resolved) ARPES with its resolution in momentum space.


The example of BiH shows that the
TRS breaking associated with the restriction to one spin species 
results in \emph{total} dichroism. Analogous effects are present in graphene with enhanced SOC.
Hence, measuring circular dichroism from a 2D system
allows to directly determine its topological property, even for a TI with
overall TRS. The key aspect is the spin resolution provided by
spin-resolved ARPES. This is in contrast to, for instance, optical
absorption, which could not distiniguish the spin species and would thus result
in zero dichroism for spin Chern insulators, which constitute the majority of
existing 2D TIs. Hence, measuring circular dichroism in spin-resolved ARPES
provides a powerful tool for the identification of TIs.

Furthermore, the extension of ARPES to the time domain
(tARPES)~\cite{schmitt_transient_2008,wang_observation_2013,sentef_examining_2013},
offers a new way of tracing and defining transient topological phenomena. For
instance, the build-up of light-induced topological
states~\cite{kitagawa_transport_2011,sentef_theory_2015,dahlhaus_magnetization_2015,claassen_all-optical_2016,hubener_creating_2017,topp_all-optical_2018,claassen_universal_2018,mciver_light-induced_2018}
should be observable with tARPES in real time. This is particularly important as
laser-heating effects typically lead to thermalization at high
temperature, where the Hall conductance is not
quantized~\cite{schuler_quench_2018,Sato2019a}. In contrast, the energy
selectivity of ARPES allows to identify the topological character
of the individual bands, thus providing a conclusive result even in
highly excited systems.

\begin{acknowledgements}
We acknowledge helpful discussion with Peizhe Tang.
Furthermore we acknowledge financial support from the Swiss National Science Foundation via NCCR MARVEL
and the European Research Council  via ERC-2015-AdG-694097 and ERC Consolidator Grant No.~724103. The Flatiron Institute is a division of the Simons Foundation. M.~S. thanks the Alexander von Humboldt
Foundation for its support with a Feodor Lynen scholarship. M.~A.~S.~acknowledges financial support by the DFG through the Emmy Noether program (SE 2558/2-1).
\end{acknowledgements}

\appendix

\section{Spin-orbit coupling effects in graphene and BiH \label{sec:soc2d}}

In this appendix we show that $S_z$ as approximate quantum number for the systems 
with SOC which we discuss in this work. To this end, we have solved the Kohn-Sham equations including the SOC (numerical details in Appendix~\ref{sec:tddft_details}), treating the Bloch wave-functions $|\Psi_{\vec{k}\alpha}\rangle$ as general spinors. 
This allows for calculating the expectation value $S_z(\vec{k})=\langle \Psi_{\vec{k}\alpha}| \hat{s}_z | \Psi_{\vec{k}\alpha} \rangle$, where $\hat{s}_z = (\hbar/2)\hat{\sigma}_z$ denotes the operator of the z-projection of the spin. 

\begin{figure}[t]
\centering
\includegraphics[width=\columnwidth]{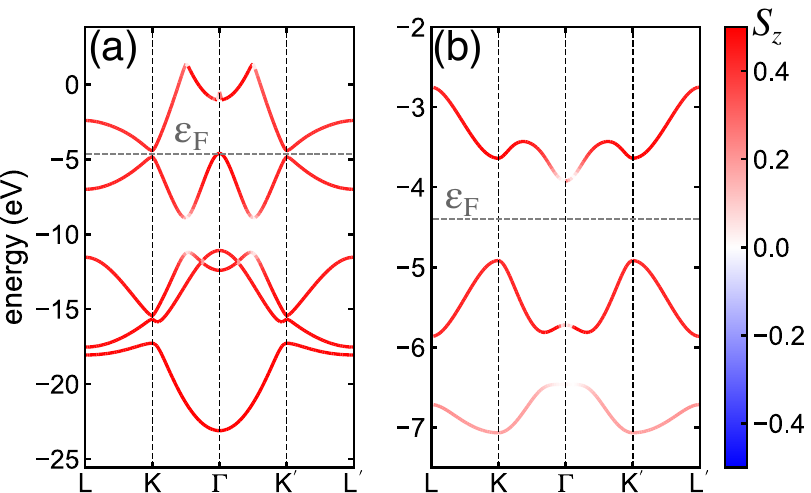}
\caption{Spin expectation value $S_z(\vec{k})$ of the predominantly spin-up bands for (a) graphene with enchanced SOC, and (b) BiH. The color bar is in units of $\hbar$.\label{fig:soc_spin_bands}}
\end{figure}

Fig.~\ref{fig:soc_spin_bands} shows $S_z(\vec{k})$ for both graphene with enhanced SOC as well as for 
BiH with full intrinsic SOC. We have focused on the bands with predominant spin-up character close to the Fermi energy. As Fig.~\ref{fig:soc_spin_bands} demonstrates, $S_z(\vec{k})$ is very close to $+\hbar/2$ in the vicinity of the K and K$^\prime$ point for the top valence band and most parts of the bottom conduction band. Hence, $S_z$ can be regarded as good quantum number, justifying the block-diagonal structure of the Hamiltonian~\eqref{eq:Ham}.

\section{wave packet picture\label{sec:wp_details}}

To understand the self-rotation and the associated orbital magnetic
moment, we employ the wave packet picture~\cite{ashcroft_solid_1976}. Let us consider a wave packet with respect to band $\alpha$ of the form
\begin{align}
  \label{eq:wp_construct}
  W_{\vec{k}\alpha}(\vec{r}) = \int\!d \vec{q}\, a(\vec{k},\vec{q})
  \psi_{\vec{q}\alpha}(\vec{r}) = \int\!d \vec{q}\, a(\vec{k},\vec{q}) e^{\iu
  \vec{q}\cdot \vec{r}} u_{\vec{q}\alpha}(\vec{r}) \ .
\end{align}
For computing ARPES matrix elements, it is convenient to introduce an analogue of cell-periodic functions
by $F_{\vec{k}\alpha}(\vec{r})=e^{-\iu \vec{k}\cdot\vec{r}}W_{\vec{k}\alpha}(\vec{r})$.
The envelope function $a(\vec{k},\vec{q})$ represents a narrow distribution
around a central wave vector $\vec{k}$; its precise functional form
does not play a role. Denoting the center of the wave packet by 
\begin{align}
  \vec{r}_c = \langle W_{\vec{k}\alpha} | \vec{r} |  W_{\vec{k}\alpha} \rangle = 
  \langle F_{\vec{k}\alpha} | \vec{r} |  F_{\vec{k}\alpha} \rangle \ ,
\end{align}
one defines~\cite{xiao_berry_2005} the angular momentum as 
\begin{align}
  \label{eq:wp_ang}
  \langle \hat{\vec{L}} \rangle = \langle W_{\vec{k}\alpha} |
  (\vec{r}-\vec{r}_c) \times \hat{\vec{p}} | W_{\vec{k}\alpha} \rangle = \langle F_{\vec{k}\alpha} |
  (\vec{r}-\vec{r}_c) \times \hat{\vec{p}} | F_{\vec{k}\alpha} \rangle\ ,
\end{align}
where $\hat{\vec{p}}$ denotes the momentum operator. The wave packet
representation of OAM~\eqref{eq:wp_ang} naturally
leads to the so-called modern theory of
magnetization~\cite{aryasetiawan_modern_2017} in the limit of
$a(\vec{k},\vec{q}) \rightarrow \delta(\vec{q}-\vec{k})$. 

\subsection{Expansion in eigenfunctions of angular momentum}

To quantify the OAM, we expand the wave packet 
$F_{\vec{k}\alpha}(\vec{r})$ onto eigenfunctions of the OAM $\hat{L}_z$:
\begin{align}
  \label{eq:wp_angbasis}
  F_{\vec{k}\alpha}(\vec{r}) = \frac{1}{\sqrt{2\pi}}\sum_m
  \mathcal{F}_{\vec{k}\alpha,m}(s,z) e^{\iu m \theta} \ .
\end{align}
Here, $\theta$ is the angle measured in the 2D plane, taking
$\vec{r}_c$ as the origin, while $s=|\vec{r}_\parallel-\vec{r}_c|$ is the
corresponding distance. Inserting the expansion~\eqref{eq:wp_angbasis}
into Eq. ~\eqref{eq:wp_ang} yields the simple expression
\begin{align}
  \langle \hat{L}_z \rangle &= \sum_m m \int^\infty_0\!d s\, s
  \int^\infty_{-\infty}\! d z\, \left| \mathcal{F}_{\vec{k}\alpha,m}(s,z)
                              \right|^2 \nonumber \\
                            &\equiv \sum_m m
                              P_{\vec{k}\alpha}(m) \ .
\end{align}
Hence, a nonzero orbital angular momentum projection in the $z$ direction
can be associated with an imbalance of the occupation of angular
momentum states $P_{\vec{k}\alpha}(m)$.

\subsection{Photoemission matrix elements}

Approximating the initial Bloch states $|\psi_{\vec{k}\alpha}\rangle$ by the 
wave packet state $|W_{\vec{k}\alpha}\rangle$ and the final states by plane waves, the dipole matrix elements are given by
\begin{align}
  \label{eq:wp_matel}
  M^{(\pm)}_\alpha(\vec{p},p_\perp) &= \int\!d\vec{r}_\parallel \!
  \int^\infty_{-\infty}\!d z\, e^{-\iu \vec{p}\cdot \vec{r}_\parallel}
  e^{-\iu p_\perp z} (x\pm \iu y) W_{\vec{k}\alpha}(\vec{r}) \nonumber \\
  &=  \frac{1}{\sqrt{2}}\int\!d\vec{r}_\parallel \!
  \int^\infty_{-\infty}\!d z\, e^{-\iu (\vec{p}-\vec{k})\cdot \vec{r}_\parallel}
  e^{-\iu p_\perp z} (x\pm \iu y) F_{\vec{k}\alpha}(\vec{r})\ .
\end{align}
Now we insert the angular-momentum
representation~\eqref{eq:wp_angbasis} and the plane-wave expansion
around $\vec{r}_c$
\begin{align*}
  e^{\iu (\vec{p}-\vec{k})\cdot\vec{r}_\parallel} = e^{\iu (\vec{p}-\vec{k})\cdot\vec{r}_c}\sum_m \iu^m J_m(|\vec{p}-\vec{k}|
  s) e^{\iu m (\theta-\theta(\vec{p},\vec{k}))} \ ,
\end{align*}
where $\theta(\vec{p},\vec{k})$ is the angle defining the direction of
$\vec{p}_\parallel-\vec{k}$, into Eq. ~\eqref{eq:wp_matel}. 
Thus, we can express the matrix elements as
\begin{align}
  \label{eq:wp_matel_cd}
&M^{(\pm)}_\alpha(\vec{p},p_\perp) = e^{\iu (\vec{p}-\vec{k})\cdot\vec{r}_c}\sum_m
  (-\iu)^{m\pm 1}
  e^{\iu (m\pm 1) \theta(\vec{p},\vec{k})}  \nonumber \\ & \times\int^\infty_0\!d s\,
  \int^\infty_{-\infty}\! d z\, e^{-\iu p_\perp z} J_{m\pm
  1}(|\vec{p}-\vec{k}|s) s^2 \mathcal{F}_{\vec{k}\alpha,m}(s,z) \ .
\end{align}
Assuming the distribution $a(\vec{k},\vec{q})$ to be sufficiently narrow, such that Bloch states are recovered,
the energy conservation implies $\vec{p}\approx \vec{k}$. As $J_m(x)\rightarrow 0$ for $x\rightarrow 0$ with $m \ne 0$, only the term with $m=0$ contributes to the sum in Eq. ~\eqref{eq:wp_matel_cd}. The dominant matrix element simplifies to
\begin{align}
  \label{eq:wp_matel_cd_simple}
  M^{(\pm)}_\alpha(\vec{k},p_\perp) = \int^\infty_0\!d s\,
  \int^\infty_{-\infty}\! d z\, e^{-\iu p_\perp z} s^2 \mathcal{F}_{\vec{k}\alpha,\mp 1}(s,z) \ .
\end{align}
This expression demonstrates 
that the asymmetry of OAM eigenstates with $m=\pm 1$ determine the circular dichroism. 

\subsection{Illustration for hBN}

\begin{figure}[t]
  \centering
  \includegraphics[width=\columnwidth]{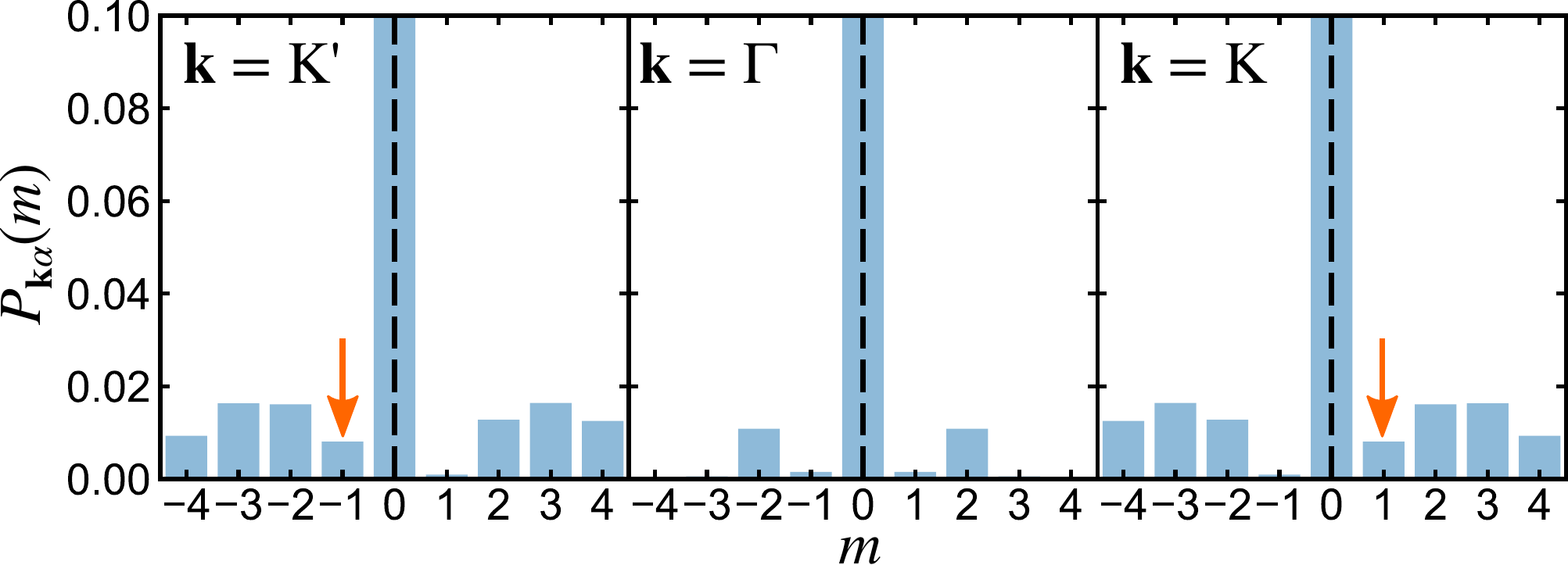}
  \caption{Weights of the projections onto OAM eigenfunctions according to Eq. ~\eqref{eq:wp_angbasis} for the TB model of hBN. The width of the
    distribution is $\Delta k=0.025$ in reduced coordinates. The arrows indicate the dominant contribution for the matrix elements~\eqref{eq:wp_matel_cd_simple}. \label{fig:hbn_GKKp_mproj}}
\end{figure}

In order to illustrate the discussion above, we have constructed
Bloch wave packets according to Eq. ~\eqref{eq:wp_construct}, choosing
a distibution function{}
$a(\vec{k},\vec{q})=a_0\exp[-(\vec{q}-\vec{k})^2/(2\Delta k^2)]$ ($a_0$ is a
normalization constant). The underlying Bloch wave functions are
constructed using the TB model for hBN.

We have computed the projection onto planar OAM eigenfunctions
(Eq. ~\eqref{eq:wp_angbasis}) and the corresponding weights
$P_{\vec{k}\alpha}(m) =$ for the valence band ($\alpha=v$),
as presented in Fig.~\eqref{fig:hbn_GKKp_mproj}. 
As Fig.~\eqref{fig:hbn_GKKp_mproj} demonstrates, the 
OAM eigenstate $m=-1$ ($m=+1$) dominates at $\vec{k}=\mathrm{K}^\prime$ ($\vec{k}=\mathrm{K}$). 
At $\vec{k}=\mathrm{K}^\prime$, we expect photoelectrons emitted by RCP light -- this is in line with Fig.~\ref{fig:hbn}. The behavior at $\vec{k}=\mathrm{K}$ is reversed.
In contrast, at $\vec{k}=\Gamma$ the weights are symmetric. Hence, vanishing dichroism is expected around the $\Gamma$-point; we have confirmed this behavior by explicitly calculating the circular dichroism within the full TB+PW model.

\section{\emph{Ab-initio} ARPES simulations: numerical details\label{sec:tddft_details}}

The evolution of the electronic structure under the effect of external fields was computed 
by propagating the Kohn-Sham (KS) equations in real space and real time within TDDFT as implemented 
in the Octopus code~\cite{Marques200360,castro_octopus:_2006,Strubbe:2015iz,noauthor_octopuswiki_nodate}. 
We solved the KS equations in the local density approximation (LDA)~\cite{Perdew:1981dv} with semi-periodic 
boundary conditions. 
For all the systems considered, we used a simulation box of 120~$a_0$ along the non-periodic dimension and the 
primitive cell on the periodic dimensions with a grid spacing of 0.36~$a_0$, and sampled the Brillouin zone 
with a 12$\times$12 {\bf k}-point grid. 
We modeled graphene with a lattice parameter of 6.202~$a_0$ and hBN with 4.762~$a_0$.  
Time and spin-resolved ARPES was calculated by recording the flux of the photoelectron current over a 
surface placed 30~$a_0$ away from the system with the t-SURFFP 
method~\cite{de_giovannini_first-principles_2017,DeGiovannini:2018do} -- 
the extension of t-SURFF~\cite{Scrinzi:2012jt,Wopperer:2017bm} to periodic systems.
All calculations were performed using fully relativistic HGH pseudopotentials~\cite{Hartwigsen:1998dk}.

\section{Tight-binding modelling\label{sec:tb_details}}

\subsection{Tight-binding representation of initial states}

Within the TB model, we approximate
the Bloch states $\psi_{\vec{k}\alpha}(\vec{r})$ by
\begin{align}
  \label{eq:wannrep}
  \psi_{\vec{k}\alpha}(\vec{r}) &=\frac{1}{\sqrt{N}}\sum_{\vec{R}}
  e^{\iu\vec{k}\cdot\vec{R}} \phi_{\vec{k}\alpha}(\vec{r}-\vec{R})
                                  \nonumber \\ &=  \frac{1}{\sqrt{N}}\sum_{\vec{R}}
  \sum_j C_{\alpha j}(\vec{k}) e^{\iu\vec{k}\cdot (\vec{R}+\vec{t}_j)}
  w_j(\vec{r}-\vec{R}) \ .
\end{align}
Here, we are
employing the convention where the phase factor
$e^{\iu\vec{k}\cdot\vec{t}_j}$ ($\vec{t}_j$ denotes the sublattice
site positions) is directly included in the definition~\eqref{eq:wannrep}
of the Bloch states. 

For all considered systems, we have constructed a nearest-neighbour
(NN) TB model and fitted the onsite and hopping energy to the
respective bandstructure of the DFT calculation. For graphene with
enhanced SOC, we have used the next-NN model
from Ref.~\cite{kane_quantum_2005} and fitted the corresponding SOC
parameter. For BiH, we used the effective TB Hamiltonian 
from Ref.~\cite{li_theoretical_2018} for the subset of $p_x$ and $p_y$ orbitals. In all cases, the bandstructure obtained by the TB models matches the DFT energies close to the K and K$^\prime$ point
very well. 

The TB Wannier orbitals are approximated as
\begin{align}
  w^\gamma_{j}(\vec{r}) = C_j \vec{u}_\gamma \cdot(\vec{r}-\vec{t})_j \exp\left[-\alpha_j
  (\vec{r}-\vec{t}_j)^2\right] \ ,
\end{align}
where $\vec{u}_\gamma$ is the unit vector in the direction $\gamma=x,y,z$.
The parameters $C_j$ and $\alpha_j$ are fitted to atomic orbitals. 

\subsection{Matrix elements}

To further simplify the analysis, we approximate the final states as
plane-waves (PW). The cell-periodic part
$\tilde{\chi}_{\vec{p},p_\perp} (\vec{r})= e^{-\iu \vec{k}\cdot
  \vec{r}} \chi _{\vec{p},p_\perp} (\vec{r})$  thus reduces to
$\tilde{\chi}_{\vec{p},p_\perp} (\vec{r})=N^{-1/2} e^{\iu p_\perp
  z}$, where $N$ is the normalization as in the Wannier
representation~\eqref{eq:wannrep}. 

Due to the periodicity of both the initial and final states, the
matrix element entering Eq. ~\eqref{eq:arpes},
\begin{align}
  \label{eq:matel_def}
  M_\alpha(\vec{p},\pp) = \langle \chi _{\vec{p},\pp} |
  \hat{\epsilon}\cdot \vec{r} |\psi_{\vec{k}\alpha} \rangle,
\end{align}
is only nonzero if
$\vec{p}=\vec{k}+\vec{G}$, where $\vec{G}$ is a reciprocal lattice
vector. Here, we focus on ARPES from the first BZ, so that $\vec{G}=0$. 

To evaluate the photoemission matrix element in the length gauge, we
employ the identity
\begin{align*}
  \vec{r} \psi_{\vec{k}\alpha}(\vec{r}) = \iu
  e^{\iu\vec{k}\cdot\vec{r}} \nabla_{\vec{k}}
  u_{\vec{k}\alpha}(\vec{r})  - \iu \nabla_{\vec{k}}
  \psi_{\vec{k}\alpha}(\vec{r}) \ ,
\end{align*}
which transforms the dipole operator into a cell-periodic
expression. Inserting the Wannier representation~\eqref{eq:wannrep}
into Eq. ~\eqref{eq:matel_def},
we find for the matrix elements
\begin{widetext}
\begin{align}
   M_\alpha(\vec{k},\pp) &= \frac{\iu}{\sqrt{N}} \sum_{\vec{R}}
                            \int\!\dd\vec{r}\,\tilde{\chi}^*_{\vec{k}\pp}(\vec{r}) \pol\cdot
                           \nabla_{\vec{k}}
                           \left[e^{-\iu\vec{k}\cdot(\vec{r}-\vec{R})}
                           \phi_{\vec{k}\alpha}(\vec{r}-\vec{R})\right]
                           -  \frac{\iu}{N} \sum_{\vec{R}}
                           \int\!\dd\vec{r}\,\chi^*_{\vec{k}\pp}(\vec{r})
                           \pol\cdot
                           \nabla_{\vec{k}}
                           \left[e^{\iu\vec{k}\cdot\vec{R}}
                           \phi_{\vec{k}\alpha}(\vec{r}-\vec{R})\right] \nn
  \\ &= \sqrt{N}\int\!\dd\vec{r}\,\chi^*_{\vec{k}\pp}(\vec{r})
                           \pol\cdot \vec{r}
       \phi_{\vec{k}\alpha}(\vec{r}) - \frac{1}{\sqrt{N}} \sum_{\vec{R}}
       \pol\cdot\vec{R} \int\!\dd\vec{r}\,\chi^*_{\vec{k}\pp}(\vec{r})
       \phi_{\vec{k}\alpha}(\vec{r}) \ . \label{eq:lenght1}
\end{align}
\end{widetext}
The derivation is analogous to Ref.~\cite{park_orbital_2012}.
The second term in Eq. ~\eqref{eq:lenght1} vanishes.

Note that the origin $\vec{r}_0$ from which the dipole is measured 
($\vec{r}\rightarrow \vec{r}-\vec{r}_0$) is arbitrary if exact
scattering states $|\chi_{\vec{p},\pp}\rangle$ are used. However,
within the PW approximation, the initial and final states are not exactly
orthogonal, which results in a slight dependence on
$\vec{r}_0$. Here, we consistently choose $\vec{r}_0 = (\vec{t}_B-\vec{t}_A)/2$,
where $j=A,B$ denotes the sublattice sites. This choice encodes as
many symmetries as possible and leads to a very good agreement of the
ARPES intensity between TDDFT and the TB approach. 

Defining the Fourier transformed Wannier orbitals by
\begin{align}
  \varphi_j(\vec{k},\pp) =  \int\!\dd\vec{r}\,
  e^{-\iu\vec{k}\cdot\vec{r}} e^{-\iu \pp z} w_j(\vec{r}) \ ,
\end{align}
the matrix elements can be expressed via
\begin{align}
  \label{eq:matrix_len}
  M_\alpha(\vec{k},\pp) = \iu \sum_j C_{\alpha
  j}(\vec{k})e^{\iu\vec{k}\cdot\vec{t}_j} \pol\cdot\left[
  \iu\nabla_{\vec{k}} -\vec{r}_0\right]
  \varphi_j(\vec{k},\pp) \ .
\end{align}

\section{Pseudospin picture\label{sec:pseudospin}}

In this appendix we demonstrate that the circular dichroism for $\pi$-conjugate
systems like graphene and hBN is directly related to the orbital
pseudospin. This provides a clear link to a topological phase
transition, which is characterized by a sign change of the pseudospin
in the BZ.

The starting point is the expression~\eqref{eq:matrix_len} for the
matrix element in the length gauge. The Fourier transformation of the
Wannier orbital centered around $\vec{t}_j$ can be conveniently
expressed as
$\varphi_j(\vec{k},\pp) = e^{-\iu \vec{k}\cdot \vec{t}_j}
\widetilde{\varphi}_j(\vec{k},\pp)$. In particular, if the Wannier
orbital $w_j(\vec{r})$ is radially symmetric around $\vec{t}_j$ and
symmetric or antisymmetric along the $z$-axis,
$\widetilde{\varphi}_j(\vec{k},\pp)$ becomes a purely real or
imaginary function.  Simplifying Eq. ~\eqref{eq:matrix_len} in this
way, we obtain
\begin{align*}
  M_\alpha(\vec{k},\pp) = \pol\cdot \sum_j C_{\alpha
  j}(\vec{k})(\vec{t}_j-\vec{r}_0 + \iu
  \nabla_{\vec{k}})
  \widetilde{\varphi}_j(\vec{k},\pp) \ .
\end{align*}
The difference of the modulus squared matrix elements upon inserting
$\pol^{(\pm)}$ yields
\begin{widetext}
\begin{align}
\label{eq:length_deltaM}
  \Delta\mathcal{M}_{\alpha}(\vec{k},\pp) &\equiv
  \left|M^{(+)}_{\alpha} (\vec{k},\pp)\right|^2 -
  \left|M^{(-)}_{\alpha} (\vec{k},\pp)\right|^2 \nn \\ &=
2\mathrm{Im}\sum_{j,l} C_{\alpha j}(\vec{k}) C^*_{\alpha l}(\vec{k})
                                                          \left[[(\vec{t}_j-\vec{r}_0)
  \widetilde{\varphi}_j(\vec{k},\pp) + \iu
  \nabla_{\vec{k}}\widetilde{\varphi}_j(\vec{k},\pp) \right] \times \left[(\vec{t}_l-\vec{r}_0)
  \widetilde{\varphi}^*_l(\vec{k},\pp) - \iu
  \nabla_{\vec{k}}\widetilde{\varphi}^*_l(\vec{k},\pp) \right] \ ,
\end{align}
where we take the $z$ component of the vector product. Further
evaluating Eq. ~\eqref{eq:length_deltaM}, the matrix element
asymmetry can be decomposed into two terms
\begin{align}
  \Delta\mathcal{M}_{\alpha}(\vec{k},\pp)  =
  \Delta\mathcal{M}^{(1)}_{\alpha}(\vec{k},\pp) +
  \Delta\mathcal{M}^{(1)}_{\alpha}(\vec{k},\pp) \ ,
\end{align} 
where
\begin{align}
  \label{eq:length_deltaM1}
  \Delta\mathcal{M}^{(1)}_{\alpha}(\vec{k},\pp) = 2\mathrm{Re}\sum_{j,l} C_{\alpha j}(\vec{k}) C^*_{\alpha l}(\vec{k})
                                                          \left[(\vec{t}_l-\vec{r}_0)
  \widetilde{\varphi}^*_l(\vec{k},\pp) \times 
   \nabla_{\vec{k}}\widetilde{\varphi}_j(\vec{k},\pp) +(\vec{t}_j-\vec{r}_0)
  \widetilde{\varphi}_j(\vec{k},\pp) \times 
   \nabla_{\vec{k}}\widetilde{\varphi}^*_l(\vec{k},\pp)  \right]
\end{align} 
and
\begin{align}
 \label{eq:length_deltaM2}
  \Delta\mathcal{M}^{(2)}_{\alpha}(\vec{k},\pp) =
  2\mathrm{Im}\sum_{j,l} C_{\alpha j}(\vec{k}) C^*_{\alpha l}(\vec{k})
\nabla_{\vec{k}}\widetilde{\varphi}_j(\vec{k},\pp)\times
  \nabla_{\vec{k}}\widetilde{\varphi}^*_l(\vec{k},\pp) \ .                                                        
\end{align}
\end{widetext}
Both the contributions~\eqref{eq:length_deltaM1} and
\eqref{eq:length_deltaM2} are important. However, assuming a radial
symmetry of the Wannier orbitals around their center renders
$\widetilde{\varphi}_j(\vec{k},\pp)$ real and, furthermore,
$\widetilde{\varphi}_j(\vec{k},\pp)=\widetilde{\varphi}_j(k,\pp)$. In
this case, $ \Delta\mathcal{M}^{(2)}_{\alpha}(\vec{k},\pp) =0$.

Let us now specialize to the two-band TB model of graphene or hBN. The atomic
p$_z$ orbitals fulfill the above requirement. Thus, we arrive at
\begin{align*}
  \Delta\mathcal{M}^{(1)}_{\alpha}(\vec{k},\pp)  =
  \frac{4}{k}\sum_{j,l}&\mathrm{Re}\left[ C_{\alpha j}(\vec{k}) C^*_{\alpha
  l}(\vec{k}) \right] \\ &\cdot \left((\vec{t}_l-\vec{r}_0)\times \vec{k}\right)
  \widetilde{\varphi}_j(k,\pp) \widetilde{\varphi}^\prime_l(k,\pp) \ .
\end{align*}
Here,
$\nabla_{\vec{k}}\widetilde{\varphi}_j(k,\pp) = (\vec{k}/k)
\widetilde{\varphi}^\prime_j(k,\pp)$ has been exploited. Furthermore,
the sublattice sites $j=\mathrm{A,B}$ are equivalent, such that
$\widetilde{\varphi}_j(k,\pp)=\widetilde{\varphi}(k,\pp)$.
Inserting
$\vec{r}_0 = (\vec{t}_B-\vec{t}_A)/2$ and 
$\vec{t}_A=0$, $\vec{t}_B= \gvec{\tau}$, the asymmetry simplifies to
\begin{align}
\label{eq:length_deltaM1_fin1}
\Delta\mathcal{M}^{(1)}_{\alpha}(\vec{k},\pp) &= \frac{2}{k}\left(|C_{\alpha
  A}(\vec{k})|^2 -|C_{\alpha
  B}(\vec{k})|^2  \right) \left[\vec{k}\cdot \gvec{\tau}\right]_z \nn \\
  &\quad \times \widetilde{\varphi}(k,\pp) \widetilde{\varphi}^\prime(k,\pp).
\end{align}
Eq. ~\eqref{eq:length_deltaM1_fin1} contains an important message:
the difference of the
sublattice site occupation, or, in other words, the pseudospin 
\begin{align}
  \label{eq:pseudoz}
  \sigma_z(\vec{k}) = |C_{\alpha
  A}(\vec{k})|^2 -|C_{\alpha
  B}(\vec{k})|^2 
\end{align}
determines the sign of the dichroism in each valley. For graphene, one
finds $\sigma_z(\vec{k}) =0$ and hence no circular dichroism is expected. 

Furthermore, a topological phase transition can be detected based on
Eq. ~\eqref{eq:length_deltaM1_fin1}. To support this statement,
let us express the generic two-band Hamiltonian by
\begin{align}
  \label{eq:hdsigma}
  \hat{h}(\vec{k}) = \vec{D}(\vec{k}) \cdot \hat{\gvec{\sigma}} \ .
\end{align}
The main difference between a topologically trivial and nontrival
system is the zero crossing of the $D_z(\vec{k})$ component. The
states (spin-up or spin-down) correspond to sublattice sites; the
Pauli matrices represent pseuspin operators, analogous to
Ref.~\cite{sentef_theory_2015}. Suppose that the second state (spin-down)
possesses a lower energy (like in hBN, where the nitrogen lattice site
has a deeper potential), corresponding to $D_z(\vec{k}) < 0$. The
eigenstate of the Hamiltonian~\eqref{eq:hdsigma} then reads
\begin{align}
  \label{eq:coeffvec1}
  \vec{C}(\vec{k}) = \frac{1}{\mathcal{N}} \begin{pmatrix}
    D_z(\vec{k}) - |\vec{D}(\vec{k})| \\ -D_x(\vec{k}) + \iu
    D_{y}(\vec{k}) \\  \end{pmatrix} \ .
\end{align}
Evaluating the pseudospin in the $z$-direction yields
\begin{align*}
  \sigma_z(\vec{k}) = \frac{2}{\mathcal{N}} D_z(\vec{k})\left(|\vec{D}(\vec{k})| - D_z(\vec{k})\right) < 0 \ .
\end{align*}

\paragraph{Trivial case---} Assuming $D_z(\vec{k}) < 0$ across the whole BZ (which yields a
trivial band insulator) then leads to \emph{opposite} dichroism at
$\mathrm{K}$ and $\mathrm{K}^\prime$. This is a direct consequence of TRS:
$\sigma_z(\vec{k}) = \sigma_z(-\vec{k})$. Thus, we find
\begin{align}
  \label{eq:deltam_bi}
  \int_{V_1} \!
  d\vec{k}\,\Delta\mathcal{M}^{(1)}_{\alpha}(\vec{k},\pp) = -\int_{V_2} \!
  d\vec{k}\,\Delta\mathcal{M}^{(1)}_{\alpha}(\vec{k},\pp) \ .
\end{align}

\paragraph{Topologically nontrivial case---}  In contrast, a
topologically nontrivial phase is chacterized by $D_z(\vec{k}) > 0$ in
some part of the BZ. One can show that the eigenvector in the vicinity of
$\mathrm{K}^\prime$ has to be chosen as
\begin{align}
  \label{eq:coeffvec2}
  \vec{C}(\vec{k}) = \frac{1}{\mathcal{N}} \begin{pmatrix}
    D_z(\vec{k}) + |\vec{D}(\vec{k})| \\ D_x(\vec{k}) + \iu
    D_{y}(\vec{k}) \\  \end{pmatrix} \ ,
\end{align}
which results in
\begin{align*}
  \sigma_z(\vec{k}) = \frac{2}{\mathcal{N}} D_z(\vec{k})\left(|\vec{D}(\vec{k})| + D_z(\vec{k})\right) > 0 \ .
\end{align*}
For this reason, the pseudospin $\sigma_z(\vec{k})$ is has the
\emph{same} sign at both $\mathrm{K}$ and $\mathrm{K}^\prime$. Therefore,
this behavior is reflected in the matrix element
asymmtry~\eqref{eq:length_deltaM1_fin1}, and thus the
relation~\eqref{eq:deltam_bi} breaks.

Hence, the following
simple criterion can be formulated: if the valley-integrated circular dichroism has the
\emph{same} sign at $\mathrm{K}$ and $\mathrm{K}^\prime$, the system represents a
Chern insulator. This 
conclusion 
is supported by the \emph{ab initio}
calculations for graphene with enhanced SOC and the discussion of the
Haldane model in the main text.

%

\end{document}